\documentclass[a4paper]{article}
\usepackage{amsmath}
\usepackage{amsfonts}
\usepackage{cleveref}
\usepackage[utf8]{inputenc}
\usepackage[shortlabels,inline]{enumitem}
\usepackage{graphicx}
\usepackage{caption}
\usepackage{subfig}
\usepackage{float}
\usepackage[numbers,sort&compress]{natbib}
\usepackage[left=2cm,right=2cm,top=2cm,bottom=2cm]{geometry}

\setlist{noitemsep} 

\DeclareMathOperator*{\argmax}{argmax}

\newcommand{\figpage}[2]{
\begin{figure}[H]
    \centering
    \includegraphics[width=1\textwidth]{figs/beta_#2_dens_#1.pdf}
    \captionsetup{font=sf}
    \caption{Precision of given localization algorithm (row) and network (column) as function of Jaccard distance $d_J$.
    Observer density $r = #1$, infection rate $\beta = #2$, GMLA uses $K_f = \sqrt{N}$. 
    Each point is an average of over $10^3$ iterations.
    Error bars indicate three standard deviations of the mean.}
    \label{fig:r_#1_beta_#2}
\end{figure}
\begin{figure}[H]
    \centering
    \includegraphics[width=1.0\textwidth]{figs/css_beta_#2_dens_#1.pdf}
    \captionsetup{font=sf}
    \caption{CCSS$_{0.75}$ of given localization algorithm (row) and network (column) as function of Jaccard distance $d_J$.
    Observer density $r = #1$, infection rate $\beta = #2$, GMLA uses $K_f = \sqrt{N}$.
    Each point is a result of $10^3$ iterations.}
    \label{fig:css_r_#1_beta_#2}
\end{figure}
\newpage
}

\title{Impact of network topology changes on information source localization}
\author{Piotr Machura, Robert Paluch}
\date{}

\begin{document}

\maketitle

\begin{abstract}
Well-established methods of locating the source of information in a complex network are usually derived with the assumption of complete and exact knowledge of network topology.
We study the performance of three such algorithms (LPTVA, GMLA and Pearson correlation algorithm) in scenarios that do not fulfill this assumption by modifying the network prior to localization.
This is done by adding superfluous new links, hiding existing ones, or reattaching links in accordance with the network's structural Hamiltonian.
We find that GMLA is highly resilient to the addition of superfluous edges, as its precision falls by more than statistical uncertainty only when the number of links is approximately doubled.
On the other hand, if the edge set is underestimated or reattachment has taken place, the performance of GMLA drops significantly.
In such a scenario the Pearson algorithm is preferable, retaining most of its performance when other simulation parameters favor localization (high density of observers, highly deterministic propagation).
It is also generally more accurate than LPTVA, as well as orders of magnitude faster.
The aforementioned differences between localization algorithms can be intuitively explained, although a need for further theoretical research is noted.
\paragraph{Keywords:}
complex network, information propagation, source localization, exponential random graphs
\end{abstract}



\section{Introduction}
The exchange of information is the main purpose of many networks found in our everyday life.
The creation of the Internet enabled its users to communicate on an unprecedented scale.
The rapid rise in the usage of social media \cite{social_media} fundamentally changed the way we view and absorb the information presented to us by acquaintances and news publishers alike.
The emergence of clickbait \cite{clickbait} promotes sensation over honesty, encouraging the usage of exaggeration and deceit in the name of profitable engagement \cite{clickbait_recognition}.
Due to the instantaneous nature of social media misinformation propagates in a flash, reaching a wide audience of users.
As a result we, the recipients, are often unaware of the true origin of viral gossip we encounter, making it difficult to judge and interpret properly.

The scientific community has taken it upon itself to solve this dilemma, by establishing methods to identify and combat the spread of misinformation \cite{misinf_spread,countering_misinf}.
In this manner, the topic of source location is of crucial importance.
Knowing where a specific piece of information came from can aid in identifying its purpose and reliability.

Among various types of localization methods, categorized by Jiang et al. \cite{other_location}, observer-based localization methods are especially popular due to their relatively low requirements.
They can accurately pinpoint the source of information using data provided by only a few members of the network, together with the knowledge of its structure.
When deriving localization methods, it is usually assumed that this structure (network \textit{topology}) is known completely and exactly.
We propose a set of scenarios, outlined in Section \ref{sec:topology_changes}, in which this assumption does not hold.
That is, the network in which the information has spread is not the same as the network used during localization.
This may be due to errors made during network reconstruction, lack of knowledge regarding the full structure, or natural evolution occurring between the time of propagation and the moment in which the network is recorded.

Examining the simulation performance of three popular source localization algorithms, outlined in Section \ref{sec:locate}, we seek to find and describe the qualitative differences between them.
Results of such simulations can potentially aid in choosing the right algorithm for a real-world situation, in which the ``correctness'' of the reconstructed network is uncertain.

\section{Basics}
We assume a simple agent-based model, in which $N$ agents are located in the vertices of a complex network, i.e. an unweighted, undirected graph $\mathcal{G} = (V, E)$.
Edges of such graphs represent a bi-directional connection through which information may propagate.

\subsection{Information spread}
To simulate the spread of information, we use the synchronous Susceptible-Infected (SI) model \cite{si_model}.
In the SI model, each node can be in one of two states: susceptible/healthy (uninformed) or infected (informed).
In the beginning, all nodes are healthy.
At some unknown point in time, $t^*$ a single randomly chosen node becomes spontaneously infected.
We will refer to this node as the \textit{source} of information, denoted $s^*$.
In every subsequent time step, infected nodes will try to infect their susceptible neighbors, with a probability of success $\beta$, called \textit{infection rate}.
The propagation ends when there are no more healthy nodes in the network.
It is assumed that the network is connected (i.e. there is a path between each pair of nodes) during the spread.
After the propagation completes and the network topology is changed (see Section \ref{sec:topology_changes}) this assumption no longer holds.

\begin{figure}
    \centering
    \includegraphics[width=0.75\textwidth]{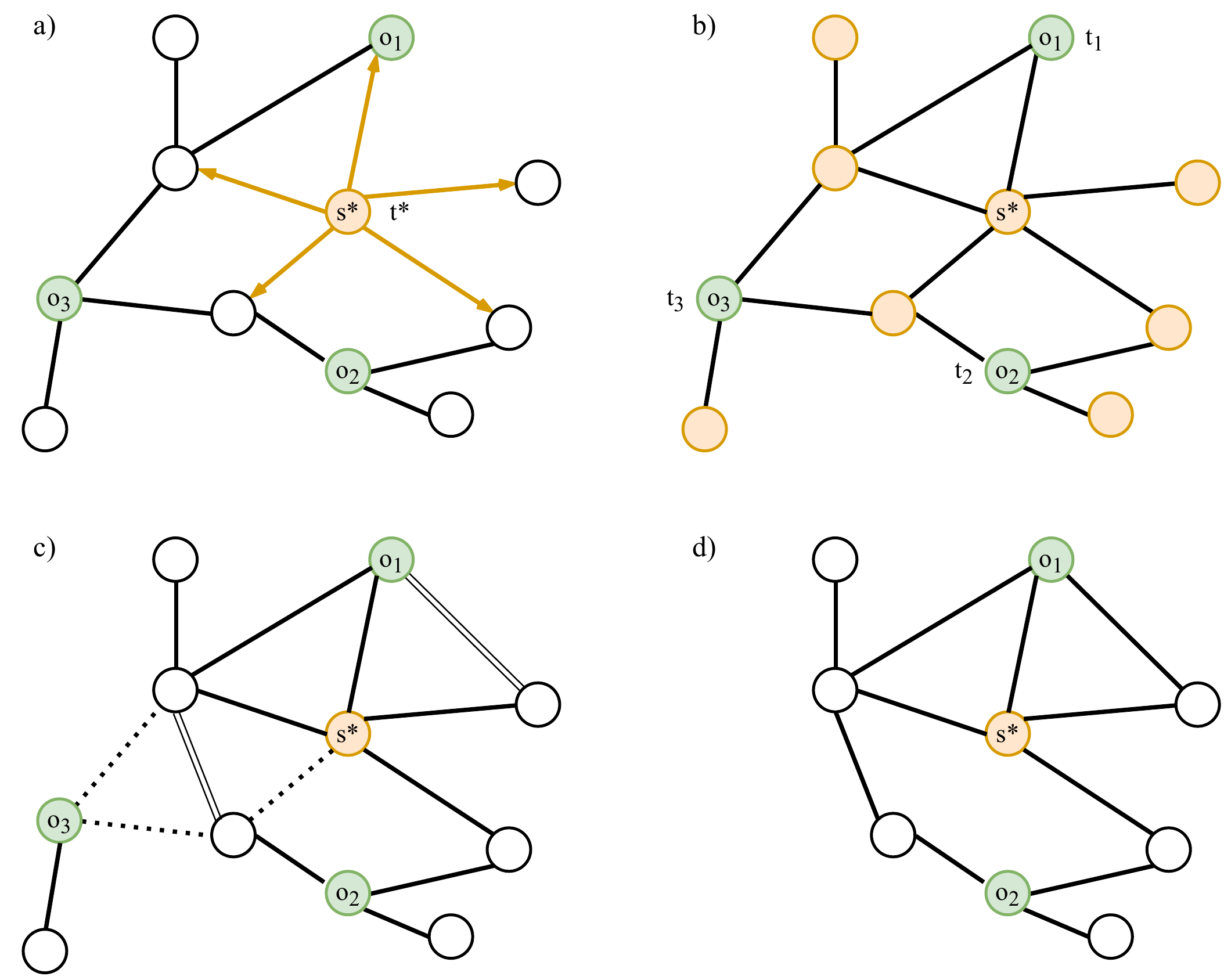}
    \caption[]{Illustration of the simulation methodology.
    \begin{enumerate*}[a)]
      \item At an unknown time $t^*$ the source $s^*$ is spontaneously infected.
      \item Information propagates through the network. 
      Infection times $t_i$ of chosen nodes (\textit{observers} $o_i$) are recorded.
      \item Network undergoes topological changes.
      Some existing edges are removed (dotted line) and some are added (double line), producing the \textit{modified} network.
      The graph may become disconnected in the process.
      \item The modified network is cropped to the largest component before the source localization algorithm is run.
    \end{enumerate*}
    }
    \label{fig:method}
\end{figure}

\subsection{Locating the source}
\label{sec:locate}

We consider \textit{observer}-based methods of source localization.
These algorithms are popular in the literature
\cite{lptv, gmla, method_comparision, observers_rumor, observers_monte_carlo}
due to their realistic assumptions, making possible real-world implementation and usage.
The observer nodes provide some limited knowledge about the spread after it has concluded.
They may represent e.g. social media moderators capable of providing insight, or individual users reporting their encounters with concerning content.

In our simulations observers form a randomly chosen subset of graph vertices $\mathcal{O} \subseteq V$, with the ratio $r = |\mathcal{O}|/|V|$ called observer \textit{density}.
It is assumed that each observer $o_i$ measures the time of its infection $t_i$.
This is notably different from assumptions made by Pinto et. al. \cite{lptv} and Xu, Teng, Zheou et al. \cite{pearson}, where both the observer infection time and the node by which the observer was infected are known.
We find that this \textit{limited} approach can still result in satisfactory algorithm performance, while significantly reducing the amount of information required to perform localization.

Source localization algorithm assigns a score to every \textit{suspect} node $s \in S \subseteq V$ (including observers), given the vector of observer infection times $[\mathbf{t}]_i = t_i$, infection rate $\beta$ and graph topology.
It is therefore equivalent to a score function $\phi(s; \mathbf{t}, \beta, \mathcal{G})$.
Intuitively, $\phi(s)$ should be associated with the posterior probability of $s$ being the true source $s^*$, that is
\begin{equation}
    \mathbb{P}(s = s^* | \mathbf{t}, \mathcal{G}) \sim f(\phi(s; \mathbf{t}, \beta, \mathcal{G})),
\end{equation}
where $f$ is any strictly increasing function.
This allows for ranking suspects by their score, with the highest scoring suspect
\begin{equation}
\label{eq:s_hat}
    \hat{s} = \argmax_{s} \phi(s; \mathbf{t}, \beta, \mathcal{G})
\end{equation}
being the most probable origin of the information.
If it is true that $\hat{s} = s^*$ then the localization is successful.
See Section \ref{sec:localization_evalutaion} for further details on evaluating localization algorithms.

\subsubsection{Maximum likelihood algorithm}
\label{sec:lptva}
The maximum likelihood algorithm has been proposed by Pinto et al in
\cite{lptv}.
We will refer to it as LPTVA (\textit{Limited Pinto-Thiran-Vitelli Algorithm}), with \textit{limited} signifying the fact that only infection times $t_i$ are used.
It is analytically derived for tree graphs and extended to general graphs in the way outlined below.

The unknown inception time $t^*$ is accounted for by choosing a reference observer $o_1$ and constructing a vector of infection delays $\mathbf{t}'$ with respect to $t_1$
\begin{equation}
    [\mathbf{t}']_i = t_{i+1} - t_1.
\end{equation}
If we assume that the graph in question is a tree $\mathcal{T}$ and let $P(v_i, v_j)$ denote the path between vertices $v_i$, $v_j$, the expected value for the $\mathbf{t}'$ vector (given $s$ is the source) can be written as
\begin{equation}
    [\boldsymbol{\mu}_s]_i = \mu \left(|P(s, o_{i+1})| - |P(s, o_1)|\right) 
    ,
\end{equation}
where $\mu$ is the mean time it takes for information to transmit through one edge.
In the case of the SI model it is equal to the mean of a geometric distribution $\mu = \frac{1}{\beta}$.
The covariance matrix $\Lambda$ of $\mathbf{t}'$ is given by
\begin{equation}
    [\Lambda]_{i,j} = \sigma^2 | P(o_{i+1}, o_1) \cap P(o_{j+1}, o_1) |
    ,
\end{equation}
where $\sigma^2$ is the variance of the time it takes for information to transmit through one edge.
In the case of SI model it is equal to variance of a geometric distribution $\sigma^2 = \frac{1 - \beta}{\beta^2}$.

Using the central limit theorem, we can postulate that $\mathbf{t}'$, as a sum of i.i.d. single-edge delays, follows a multivariate normal distribution with mean $\boldsymbol{\mu}_s$ and covariance matrix $\Lambda$.
Taking a logarithm of this distribution and stripping away $s$-independent normalization factors we arrive at the score function
\begin{equation}
    \label{eq:LPTVA_score}
    \phi_{\text{LPTVA}}(s) = -(\mathbf{t}' - \boldsymbol{\mu}_s)^T\Lambda^{-1}(\mathbf{t}' - \boldsymbol{\mu}_s) - \ln |\Lambda|
    ,
\end{equation}
which can be calculated for all nodes in the network.

If the graph in question is not a tree, the algorithm can be adjusted by calculating the score given by \eqref{eq:LPTVA_score} with respect to the breadth-first search (BFS) tree $\mathcal{T}^{(s)}_{BFS}$ rooted at $s$.
Such modification is equivalent to the assumption that the information propagates along the shortest paths connecting the source with observers.
This holds exactly when $\beta = 1$ and contributes to significantly reduced algorithm performance for non-tree graphs with $\beta$ significantly lower than 1.

\subsubsection{Gradient maximum likelihood algorithm}
\label{sec:gmla}
The LPTV algorithm, while well established and highly analytical, suffers from significant computation speed issues.
Matrix inversion in equation \eqref{eq:LPTVA_score} must be performed for each suspect, rendering the method effectively unusable for networks with a large number of nodes.

The gradient maximum likelihood algorithm (GMLA) \cite{gmla} attempts to rectify this problem by prioritizing the earliest observers and utilizing a gradient-like selection of suspects.
As a consequence, unlike in the case of LPTVA and Pearson correlation algorithm, the set of suspects is generally only a small subset of all nodes in the network.
All other nodes $\widetilde{v} \in (V \setminus S)$ are assigned the same score, effectively equal $-\infty$.
While the score function used to rank suspects in GMLA is the same as \eqref{eq:LPTVA_score}, the choice of suspects is much more sophisticated.

Firstly, only $K_f$ earliest-infected observers are used in the calculations.
The performance of the algorithm varies depending on the exact value of $K_f$, with the optimum usually estimated for each network model on a case-by-case basis.
In our calculations we have used $K_f = \sqrt{N}$ for all considered networks, which is an acceptable rule of thumb \cite{gmla}.

Secondly, the score function is not calculated for all $N$ vertices.
Instead, it is first calculated for the neighbors $S^{(1)}$ of the earliest observer $o_1$.
Next, the score is calculated for the previously unchecked neighbors of the highest scoring $\hat{s}^{(1)}$ from $S^{(1)}$, creating another set of suspects $S^{(2)}$, from which the highest scoring $\hat{s}^{(2)}$ is chosen.
This process is repeated until there is no node in $S^{(n)}$ (neighbors of $\hat{s}^{(n-1)}$ not checked in previous steps) with score higher than $\hat{s}^{(n-1)}$.
In such case, the algorithm stops and returns scores calculated for all members of 
\begin{equation}
    S = \bigcup\limits_{i = 1}^n S^{(i)}
    .
\end{equation}

\begin{figure}[htb]
    \centering
    \includegraphics[width=0.35\columnwidth]{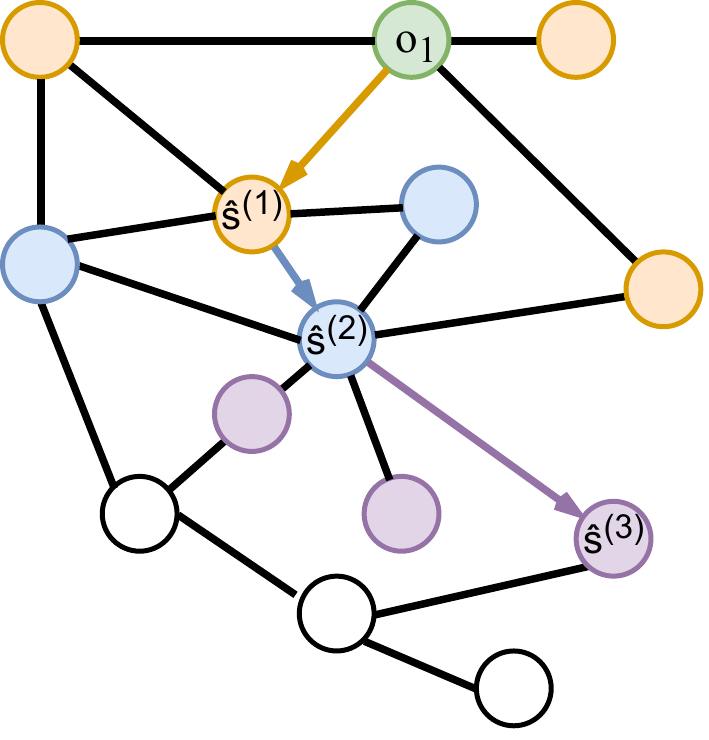}
    \caption{Node selection in GMLA.
    Initially, the neighbors of the earliest informed neighbor $o_1$ are checked ($S^{(1)}$, colored orange).
    Highest scoring vertex $\hat{s}^{(1)}$ from this set is chosen, and its neighbors $S^{(2)}$ are checked (colored blue).
    Note that $o_1$ is included in $S^{(2)}$.
    Next, neighbors $S^{(3)}$ of $\hat{s}^{(2)}$ (colored purple) are checked, among which highest scoring vertex $\hat{s}^{(3)}$ can be found.
    Since the score of $\hat{s}^{(3)}$ is lower than that of  $\hat{s}^{(2)}$, the algorithm terminates with $\hat{s} = \hat{s}^{(2)}$.
    The final suspect set consist of all colored nodes $S = S^{(3)} \cup S^{(2)} \cup S^{(1)}$.
    White vertexes remain unchecked and are placed at the bottom of the ranking, at position $N+1$, where $N$ is the size of the network.}
    \label{fig:gmla_diagram}
\end{figure}

Empirical studies suggest that the total number of suspects roughly follows $|S| \sim \hat{k} \log N$, where $\hat{k}$ is the average degree in the network.
Fig. \ref{fig:gmla_diagram} illustrates this gradient-like selection process.

GMLA ensures that nodes near high-scoring suspects are checked, while those far away are ignored, significantly reducing computation time.
Using only $K_f$ first observers also proves to be highly beneficial, bringing the overall localization capability of GMLA in line with, or even exceeding, that of LPTVA
\cite{method_comparision}.

\subsubsection{Pearson correlation algorithm}
\label{sec:pearson}
Proposed by  Xu, Teng, Zheou et al. \cite{pearson}, the correlation algorithm stands out as an extremely straightforward, yet highly performant method.
If we assume that the information propagates along the shortest path connecting the source to observers (which is also an assumption required for extending LPTVA to general graphs), then it should be intuitively obvious that the higher the distance from an observer to the source, the longer it takes for this observer to become infected.

More precisely, in the case of true source $s^*$, the vector of infection times $\mathbf{t}$ should be correlated with the vector of distances between the source and the observers
\begin{equation}
    [\mathbf{d}_{s^*}]_i = d(s^*, o_i),
\end{equation}
where distance $d(v_i, v_j)$ is defined as the length of shortest path connecting vertices $v_i$ and $v_j$.
The score, calculated for all suspects, is therefore given by the Pearson correlation coefficient between $\mathbf{t}$ and distances to the suspect $\mathbf{d}_s$
\begin{equation}
    \phi_{\text{Pearson}}(s) = \rho_{\text{Pearson}}(\mathbf{t}, \mathbf{d}_s), 
\end{equation}
which should be the highest for the true source $s = s^*$.

\subsubsection{Evaluation measures}
\label{sec:localization_evalutaion}
Algorithms outlined in Section \ref{sec:locate} assign a score to every node from the suspect set $S$.
Sorting suspects by the score in descending order produces a \textit{ranking}, with the highest scoring suspect in the first place, satisfying \eqref{eq:s_hat}.

The main measure of algorithm performance chosen for our purposes is its precision, defined as
\begin{equation}
    \text{Precision} = \frac{\text{TP}}{\text{TP} + \text{FP}}
    ,
\end{equation}
where TP is the number of true positives, or correctly located sources (1 if $\phi(s^*) = \phi(\hat{s})$, 0 otherwise) and FP is the number of false positives (all nodes $s \neq s^*$ such that $\phi(s) = \phi(\hat{s})$).
If there are no ties, then precision is a binary measure, equal to 1 if the source is at the top of the ranking and 0 otherwise.
If $s^*$ is tied with $n$ other nodes at the top of the ranking, then precision is equal to $1/n$.

The second evaluation measure chosen for the purpose of this article is the credible set size at confidence level $\alpha$, denoted $\text{CSS}_\alpha$.
Intuitively it can be understood as the size of the smallest set of top-scoring nodes containing the source with probability $\alpha$.
It is also equivalent to the $\alpha$-quantile of the rank assigned to the source by the localization algorithm.

\section{Changing network topology}
\label{sec:topology_changes}
Localization derivations outlined in Section \ref{sec:locate} assumes that the \textit{initial} network $\mathcal{G} = (V, E)$, in which the information propagates, is exactly the same as the graph used by localization algorithms $\mathcal{G}' = (V', E')$.
Such assumption, while essential for analytical derivation, is not easily satisfied in practice.
We examine a number of scenarios, for which $V = V'$ and $E \neq E'$.
That is, the set of edges of the initial network $\mathcal{G}$ is different than the set of edges of the \textit{modified} network $\mathcal{G}'$, visible to the localization algorithms.

The difference between edge sets may be due to insufficient or inaccurate information about the network's structure.
A simple interpretation of such scenario is described in Section \ref{sec:added_hidden}.
Another possible reason could be the natural changes (\textit{topological fluctuations}) taking place between the time of propagation and the moment of localization, resulting from the non-static nature of a real-world network.
In such case graphs $\mathcal{G}$ and $\mathcal{G}'$ should possess the same (or highly similar) macroscopic properties, but differ on an edge-by-edge basis.
We model this with exponential random graphs \cite{exponential_random_graphs}, using a structural Hamiltonian approach described in Section \ref{sec:fluctuations}.

The general simulation procedure, illustrated in Fig. \ref{fig:method}, can be summarized as follows:
\begin{enumerate}
    \item Create a connected network of the desired model.
    \item Designate $r \cdot N$ random nodes as observers.
    \item Propagate information according to the synchronous SI model, recording infection times $t_i$.
    \item Modify the network by adding random edges/removing random edges/performing Metropolis steps.
    \item Locate the information source using all three location algorithms. If the network became disconnected, only the largest component is searched.
\end{enumerate}

\subsection{Graph dissimilarity measure}
To compare different types of topology changes an appropriate dissimilarity measure is required.
We use Jaccard distance between $E$ and $E'$, defined as
\begin{equation}
    d_J(E, E') = 1 - J(E, E') = 1 - \frac{|E \cap E'|}{|E \cup E'|}
\end{equation}
as the free variable in most of our simulations
Other measures, namely Hamming distance, Sørensen-Dice coefficient and set overlap were considered.
Jaccard distance has been selected because it is normalized, satisfies the triangle inequality, and has an intuitive interpretation in all studied scenarios.

\subsection{Adding/hiding links}
\label{sec:added_hidden}
In this scenario the number of edges in $\mathcal{G}'$ is over- or underestimated.
In the first case a fraction of \textit{superfluous} edges $s_e$ is randomly added to the network, such that
\begin{equation}
    |E'| = (1 + s_e) |E|, \hspace{5pt} E \subseteq E'
    .
\end{equation}
The Jaccard distance is thus
\begin{equation}
    d_J(s_e) = 1 - \frac{|E|}{(1+ s_e) |E|}
    = \frac{s_e}{1 + s_e}
    ,
\end{equation}
with the inverse relationship
\begin{equation}
    s_e(d_J) = \frac{d_J}{1 - d_J}
    .
\end{equation}

Similarly, if instead of adding superfluous links we instead randomly \textit{hide} $h_e$ of them, we arrive at 
\begin{equation}
    |E'| = (1 - h_e) |E|, \hspace{5pt} E' \subseteq E,
\end{equation}
with Jaccard distance
\begin{equation}
    d_J(h_e) = 1 - \frac{(1 - h_e) |E|}{|E|} = h_e
    .
\end{equation}

It should be noted that removing edges may result in component isolation.
While the initial network is guaranteed to be connected, this is no longer the case for the modified graph.
Networks studied by us (see Section \ref{sec:results}) undergo a reversed process of percolation \cite{percolation}, while initially remaining far from the percolation threshold due to their relatively high average degree.
As a consequence, removing links means that the largest component still contains an overwhelming majority of nodes.
It is therefore the only candidate for localization, while other components consist mostly of single vertices or isolated pairs.

Further investigation of the impact of component isolation is discussed in appendix \ref{sec:isolation}.
For the purposes of our simulations we assume that, if the network splits into several components, we look for the source in the largest one.

\subsection{Topological fluctuations}
\label{sec:fluctuations}
In Section \ref{sec:added_hidden} we have described an extremely simplistic approach to topological changes, in which a number of random edges are added or removed.
We wish to study a combination of these processes, both adding and removing links.
Instead of doing it randomly (which would, in the limiting case, produce an Erdős–Rényi random graph) we consider a scheme in which some macroscopic properties of the network are preserved.
We achieve this goal using the framework of exponential random graphs.
The specific approach outlined below draws from works published by Newman et al. \cite{struct_hamiltonian} and Hołyst et al. \cite{fluctuation_dissipation}.

\subsubsection{Structural Hamiltonian}

Let $H(\mathcal{G}; \{\theta_i\})$, called \textit{structural Hamiltonian}, be a scalar function of the graphs structure and a set of parameters $\theta_i$, defined as a linear combination
\begin{equation}
\label{eq:hamiltonia_general}
    H(\mathcal{G}) = \sum\limits_i \theta_i m_i(\mathcal{G})
    ,
\end{equation}
where $m_i(\mathcal{G})$ represent any chosen properties of the graph.
If coefficients $\theta_i$ are chosen such that the probability of encountering any given graph satisfies
\begin{equation}
\label{eq:exponential_probability}
    \mathbb{P}(\mathcal{G}) \sim \exp \left(H(\mathcal{G})\right) 
    ,
\end{equation}
then the set of all possible graphs $\Omega_{\mathcal{G}}$, together with probability distribution \eqref{eq:exponential_probability}, constitutes a canonical ensemble.

We consider the case of \textit{configuration model}, for which the properties $m_i$ are individual vertex degrees $k_i$. Equation \eqref{eq:hamiltonia_general} becomes
\begin{equation}
    H(\mathcal{G}) = \sum_{i = 1}^N \theta_i k_i
    ,
\end{equation}
where $\theta_i$ must be chosen such that \eqref{eq:exponential_probability} is satisfied.
It can be shown
\cite{struct_hamiltonian}
that these coefficients are given by a system of $N$ non-linear equations
\begin{equation}
    \label{eq:nonlinear_system}
    \langle k_i \rangle =  \sum\limits_{j = 1}^N p_{ij} = \sum\limits_{j = 1}^N \frac{1}{1 + \exp(\theta_i + \theta_j)}
    ,
\end{equation}
where $p_{ij}$ is the probability of $v_i$ and $v_j$ being linked, $\langle k_i \rangle$ is the expected degree of $i$-th vertex averaging over the ensemble.
The canonical ensemble is thus defined by the expected sequence of degrees found in the network.

In the case of Erdős-Rényi (ER) model, it is easily seen that
\begin{equation}
    \theta_i = \theta = \frac{1}{2}\ln\left(\frac{ 1 - p }{p}\right)
    ,
\end{equation}
gives $\langle k_i \rangle = pN = \langle k \rangle$, where $p$ is the probability of any two nodes being linked.
For Barabási-Albert (BA) network, the degree of each vertex can be averaged over a sufficiently large number of realizations, obtaining $\langle k_i \rangle$.
Similarly, when considering real networks, the actual degree of each vertex can be designated as $\langle k_i \rangle$.
Equations \eqref{eq:nonlinear_system} can then be solved numerically for the BA model and real networks, obtaining a set of $\theta_i$ coefficients for each studied network type.

\subsubsection{Metropolis algorithm}
A common method of generating sequences of elements from a canonical ensemble is the Metropolis algorithm.
In each step, a completely random change in the current state (network structure) is performed.
Change of state is then accepted based on the resulting change in the Hamiltonian:
\begin{itemize}
    \item[--] with probability 1 if $\Delta H \geq 0$;
    \item[--] with probability $e^{\Delta H}$ if $\Delta H < 0$.
\end{itemize}
For our purposes, a change in the state consists of choosing a random pair of vertices $(v_i, v_j)$ and adding an edge $e_{ij}$ if it does not exist, or removing it if it does.
The resulting change in Hamiltonian is 
\begin{equation}
    \Delta H  = \begin{cases}
    +(\theta_i + \theta_j) & \text{when adding new } e_{ij},\\
    -(\theta_i + \theta_j) & \text{when removing existing } e_{ij}
    .
    \end{cases}
\end{equation}
The link addition/removal is then accepted in accordance with the scheme described above.

Performing a number of Metropolis steps is therefore akin to link re-attachment, with additional constraints ensuring the degree sequence is conserved when averaged over the ensemble.
The modified network produced in this manner will have edges different than $E$, but both degree sequence and (by consequence) degree distribution will be preserved.
The changes found in $\mathcal{G}'$ are therefore probable within the ensemble based on $\mathcal{G}$, with exponential probability \eqref{eq:exponential_probability}.
We interpret this as the initial network undergoing \textit{topological fluctuations}, akin to energy fluctuations of a thermodynamic system remaining in thermal equilibrium with a heat bath.

Unlike adding superfluous/hiding edges, the relationship between the number of Metropolis steps and Jaccard distance is non-trivial and network-dependent.
In our simulations, an experimental estimation of the number of steps needed to achieve $d_J = \{0, 0.1, ..., 0.7\}$ has been used.

It should be noted that the structural Hamiltonian is just one possible approach to the problem of producing a modified network with comparable macroscopic properties, but a differing set of edges.
We have chosen the exponential random graph model because of its straightforward interpretation, as well as the possibility of conserving properties other than degree sequence \cite{strauss}.

\section{Results}
\label{sec:results}
The procedure outlined in Section \ref{sec:topology_changes} has been performed for the following synthetic networks:
\begin{itemize}
    \item[--] Erdős–Rényi (ER) network with size $N = 1000$ and average degree $\langle k \rangle = 8$;
    \item[--] Barabási-Albert (BA) network with size $N = 1000$ and parameters $m_0 = m =4$.
\end{itemize}
Additionally, the following real networks have been studied:
\begin{itemize}
    \item[--] 
    Copenhagen Network Study social media friendship (\textit{,,Facebook''}) network \cite{facebook_net}.
    Vertices represent a participant of the network study, with edges indicating friendship on the Facebook social network.
    \item[--]
    University Rovira i Virgilii e-mail communication (\textit{,,Email''}) network \cite{email_net}.
    Vertices represent members of e-mail communication at the Rovira i Virgili University in Spain.
    Existence of an edge $e_{ij}$ indicates at least one e-mail message sent either from $i$ to $j$, or from $j$ to $i$.
    \item[--]
    University of California, Irvine (\textit{,,California''}) online community network \cite{california_net}, transformed into undirected graph similar to Spinelli et al \cite{california_net_transformation}.
    Vertices represent users of an online community, with edge $e_{ij}$ indicating at least two messages were sent: one from $i$ to $j$ and one from $j$ to $i$.
    The network has been trimmed such that the minimum degree is 2.
\end{itemize}
The basic properties of real networks described above are summarized in Table \ref{tab:real_properties}.
\begin{table}[H]
    \centering
    \renewcommand{\arraystretch}{1.2}
    \begin{tabular}{lccccccc}
         \textbf{Network} &  $N$ & $\langle k \rangle$ & $k_{max}$ & $\langle d \rangle$ & $d_{max}$ & $c$ \\
         \hline
         Facebook & 800 & 16.1 & 101 &  3.0 & 7 & 0.24 \\
         Email & 1133 & 9.6 & 71 & 3.5 & 5 & 0.17 \\
         California & 1020 & 12.2 & 110 & 3.0 & 8 & 0.05 \\
         \hline
    \end{tabular}
    \caption{Basic properties of real networks used in simulations.
    $\langle d \rangle$ denotes the average length of the shortest path between two nodes, $d_{max}$ is the network's diameter, $c$ is the global clustering coefficient. }
    \label{tab:real_properties}
\end{table}

\begin{figure}[h]
    \centering
    \includegraphics[width=\textwidth]{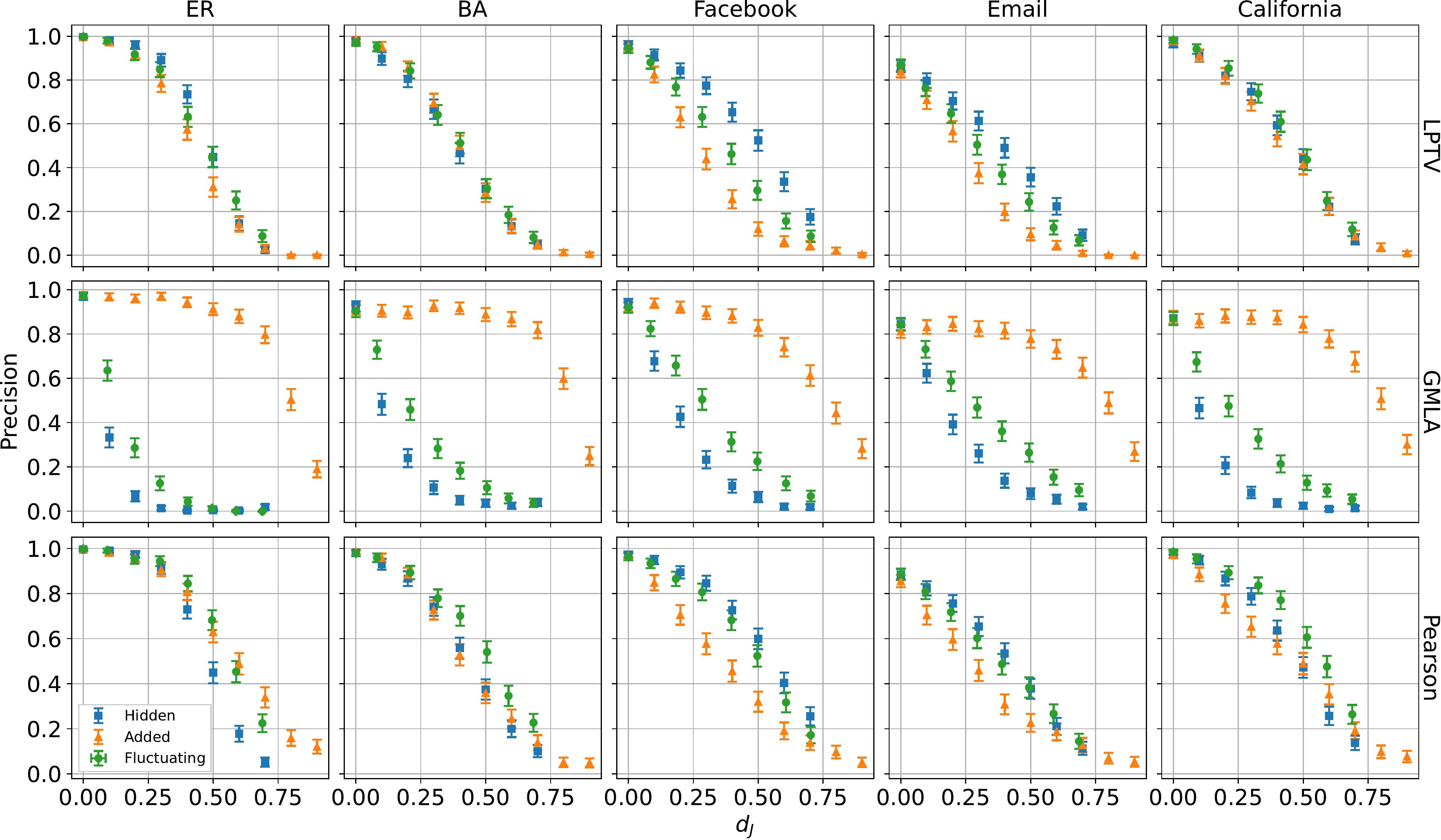}
    \caption{Precision of given localization algorithm (row) and network (column) as function of Jaccard distance $d_J$.
    Observer density $r = 0.3$, infection rate $\beta = 0.95$, GMLA uses $K_f = \sqrt{N}$. 
    Each point is an average of over $10^3$ iterations.
    Error bars indicate three standard deviations of the mean.
    }
    \label{fig:results_prec}
\end{figure}

\begin{figure}[h]
    \centering
    \includegraphics[width=\textwidth]{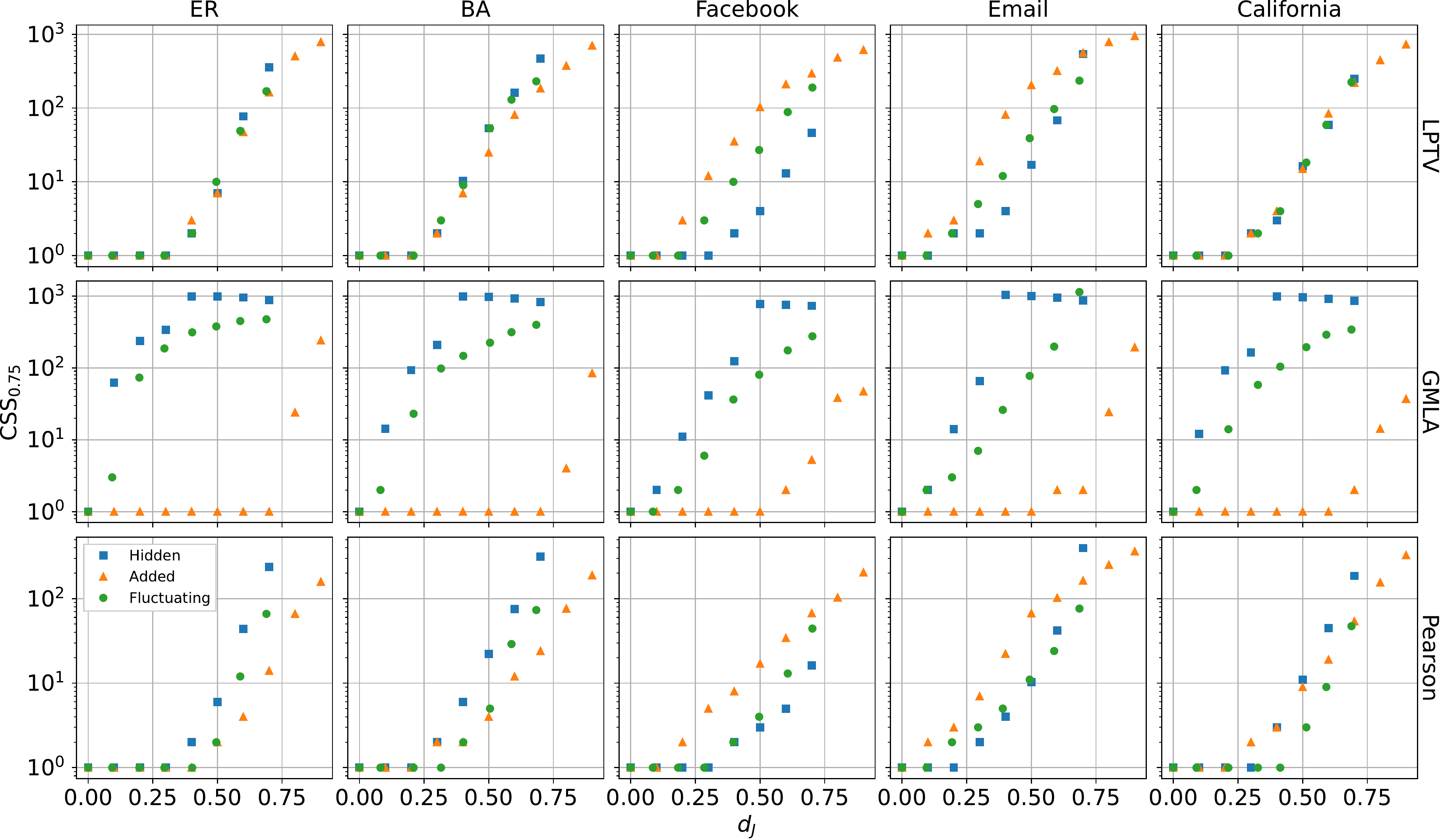}
    \caption{CSS$_{0.75}$ of given localization algorithm (row) and network (column) as function of Jaccard distance $d_J$.
    Observer density $r = 0.3$, infection rate $\beta = 0.95$, GMLA uses $K_f = \sqrt{N}$.
    Each point is a result of $10^3$ iterations.
    }
    \label{fig:results_css}
\end{figure}

Quantitative results for parameters $\beta = 0.95$, $r = 3$
are displayed in Fig. \ref{fig:results_prec} (precision) and Fig. \ref{fig:results_css} ($\text{CSS}_{0.75}$).


Upon examining Fig. \ref{fig:results_prec} we immediately notice an obvious difference in the performance of GMLA, compared to other algorithms.
It is much more resistant to the addition of superfluous edges than LPTVA, with which it shares the score function.
We observe a drop in precision exceeding statistical uncertainty only when the number of edges is approximately doubled ($d_j = 0.5$).
The same can be observed in Fig. \ref{fig:results_css}, where CSS$_{0.75}$ remains fixed at 1 for $d_J < 0.75$ for the artificial networks.
On the contrary, when hiding links, the precision of GMLA falls drastically, in some cases reaching near 0 for $d_J \geq 0.3$ (ER network).
While increasing the average vertex degree translates to a larger number of suspects being checked during the gradient-like selection process, decreasing the average vertex degree means fewer nodes are checked, reducing the probability of encountering $\hat{s}$.
This could be a sufficient explanation for the difference observed between superfluous and hidden links but does not explain the fact that topological fluctuations produce a similarly drastic, albeit consistently lesser, decrease in performance.

Notice that GMLA implicitly assumes that highly-scoring nodes should neighbor each other.
This remains true when new edges are added, but is no longer the case when hiding links or re-attaching links using the Metropolis algorithm -- the gradient is ``broken''.
Breaking the gradient should therefore be responsible for the large performance drop observed for the latter two cases, with a reduction in average vertex degree resulting in an additional difference between them.
Additionally, when considering only $K_f$ earliest observers, the paths found within the breadth-first search tree are unlikely to change when adding edges, since their length is already short (on the order of single edges).

Moreover, GMLA utilizes significantly fewer observers, which is used to filter noisy information.
In our simulations only $\sqrt{N} \approx 32$ earliest observers, out of $r \cdot N \approx 300$, are used.
This furthers the performance gap between Pearson/LPTVA and GMLA when links are hidden or re-attached, but notably does not seem to play a significant role when superfluous links are added.

Turning our attention to the Pearson algorithm, for real networks we notice a general trend of link addition having a greater impact on precision loss, with topological fluctuations placed in the middle.
For synthetic networks, the trend is reversed, with link hiding being the worse scenario, although precision differs significantly only in the case of a high $d_J$ ER network.
Difference in performance for both synthetic networks is clearly visible in the CSS$_{0.75}$ plot.
It shows a significant disparity of approximately an order of magnitude where $d_J \geq 0.6$.

Examining LPTVA performance, the observed difference between modification scenarios is notably almost indistinguishable in the case of BA and California networks.
This observation remains true for both precision and CSS$_{0.75}$.
Other networks, where the difference exceeds statistical uncertainty, favor link hiding.
Link addition has the least impact on localization, with fluctuations situated between them.
Notice that, while general curve characteristics are similar between the two, LPTVA precision is generally on par or lower than that of Pearson, together with higher CSS$_{0.75}$.
The significantly higher LPTVA execution time, arising from increased computational complexity (see \cite[Table 1]{method_comparision}), must also be taken into account.

Regardless of the localization algorithm used, some general observations should be noted.
For both Pearson and LPTV algorithms, the precision plots are generally S-shaped, sometimes displaying a degree of resilience to topological changes.
Specifically, in the case of ER network, the precision of the Pearson algorithm falls by less than the statistical uncertainty for $d_J \leq 0.3$, which translates to removing $30\%$ of existing edges or adding $\approx 43\%$ new ones.
The ER topology generally favors localization when compared to other studied networks.
Note that e.g. baseline $d_J = 0$ gives ER precision $\approx 1$ and Email precision $\approx 0.8$ for all localization algorithms.

\section{Conclusion}
\label{sec:conclusion}
In this paper, we study the relationship between the performance of three popular, observer-based algorithms and topological changes introduced in the network between the moment of propagation and localization.
We separately consider three categories of aforementioned topological modifications: hiding existing links, introducing superfluous links, and re-attaching links in accordance with structural Hamiltonian.

Considering the results presented in Section \ref{sec:results}, we propose that, when the modified network edge set is suspected to be overestimated, GMLA should be used due to its high resilience to superfluous edges.
On the contrary, when the modified network is suspected to be less dense than the initial one or undergoes topological fluctuations, GMLA should be avoided.
In such cases, the Pearson algorithm provides precision similar or higher compared to LPTVA, together with an overall lower CSS and orders of magnitude faster execution time

In some studied network-algorithm-modification combinations a degree of resilience to topological changes can be observed.
Upon examining the performance for other values of $r$ and $\beta$ (see Appendix \ref{sec:suplemental}) we note that it is most noticeable when simulation parameters favor localization (high $r$, high $\beta$, correctly chosen algorithm) and disappears when imposed conditions are worsened (low $r$, low $\beta$, incorrectly chosen algorithm).
This observation supports an intuitive idea of a ``stock'' of localization performance, which may be depleted by an unfavorable condition and replenished by a favorable one.
In this manner it is possible to ``overstock'' on precision by e.g. increasing the observer density or choosing an appropriate algorithm, thus achieving a level of resilience to topological changes. 
This idea is further supported by the fact that resilience is most clearly observed in the ER network, which by itself favors localization when compared to e.g. the Email network.

It should be noted that the overall algorithm performance is highly dependent both on the network and the remaining simulation parameters (see Appendix \ref{sec:suplemental}).
Results presented by us suggest a need for a theoretical framework, capable of rigorously explaining observed phenomena.

Furthermore, studied networks undergo percolation, as described in Appendix \ref{sec:isolation}, and as such do not split into multiple large components upon link removal.
Only one component large enough to contain an acceptable number of observers emerges.
If one were to consider scenarios where this is not the case, e.g. because of the network structure or particular link hiding scheme, then an argument can be made for a number of different approaches.
One may, for instance, look for the source in the \textit{earliest informed} component instead of the largest one, or decide to consider all sufficiently large components and try to merge their respective results.
The possibilities mentioned above warrant additional research, both theoretical and experimental.

\appendix
\section{Impact of component isolation}
\label{sec:isolation}
\begin{figure}[htb]
    \centering
    \includegraphics[width=0.4\columnwidth]{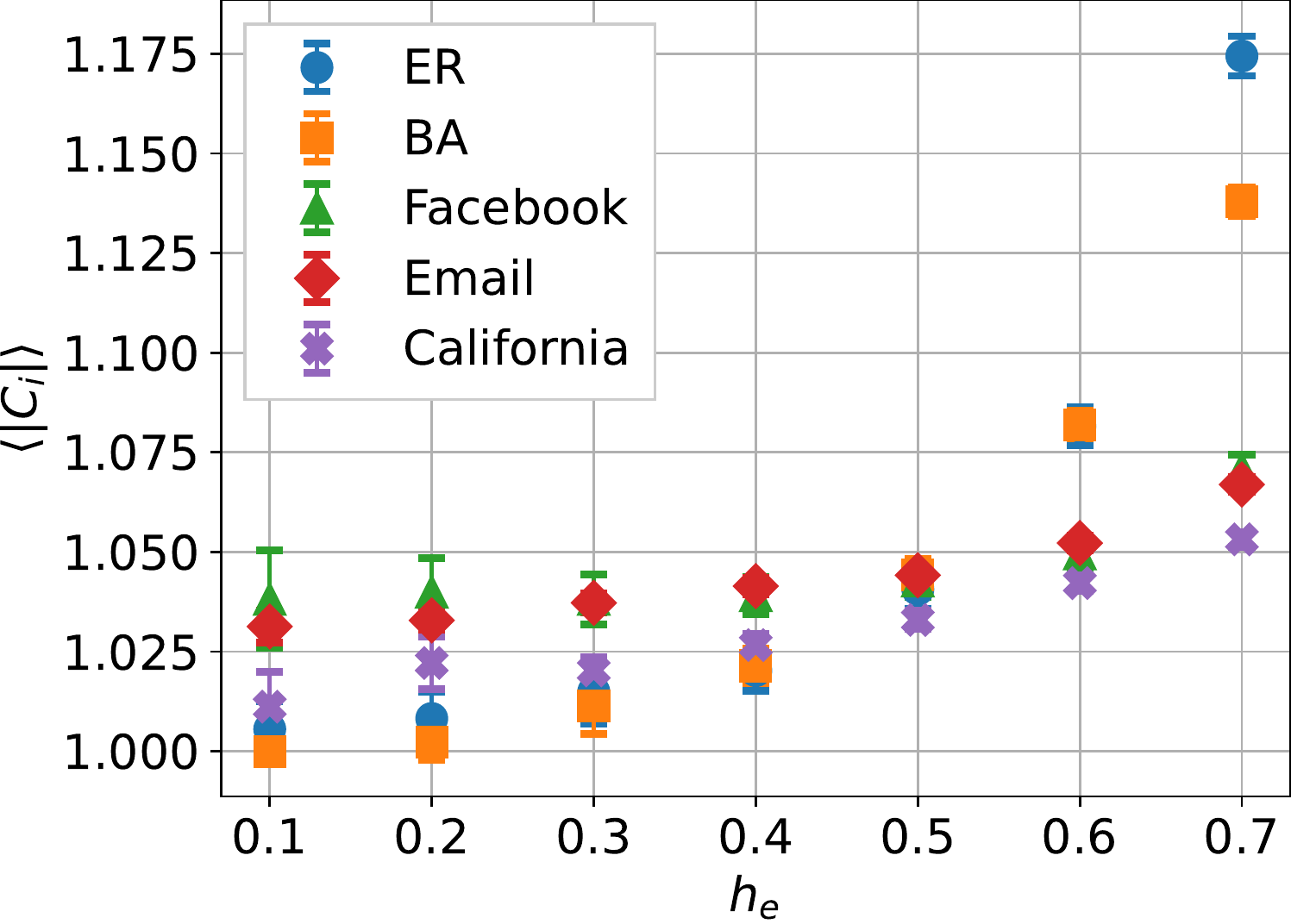}
    \caption{Average size of network clusters, excluding the largest one.
    Each point is an average over $10^3$ iterations.
    Error bars indicate three standard deviations of the mean.
    }
    \label{fig:other_components}
\end{figure}

As mentioned in Section \ref{sec:added_hidden}, hiding random edges may result in component isolation.
Let $C^*$ be the largest connected component present in the network (also called \textit{percolation cluster}), with $\{ C_i \}$ denoting the set of components other than the largest one.
If the graph became disconnected as a result of removing random edges, then by necessity $C^* \neq V$ and $\{ C_i \}$ is not empty.
Plotting the average size of $\langle |C_i| \rangle$ as a function of Jaccard distance $h_e = d_J$ (see Fig. \ref{fig:other_components}) we notice that these components are mostly single vertices.

\begin{figure}[htb]
    \centering
    \includegraphics[width=0.4\columnwidth]{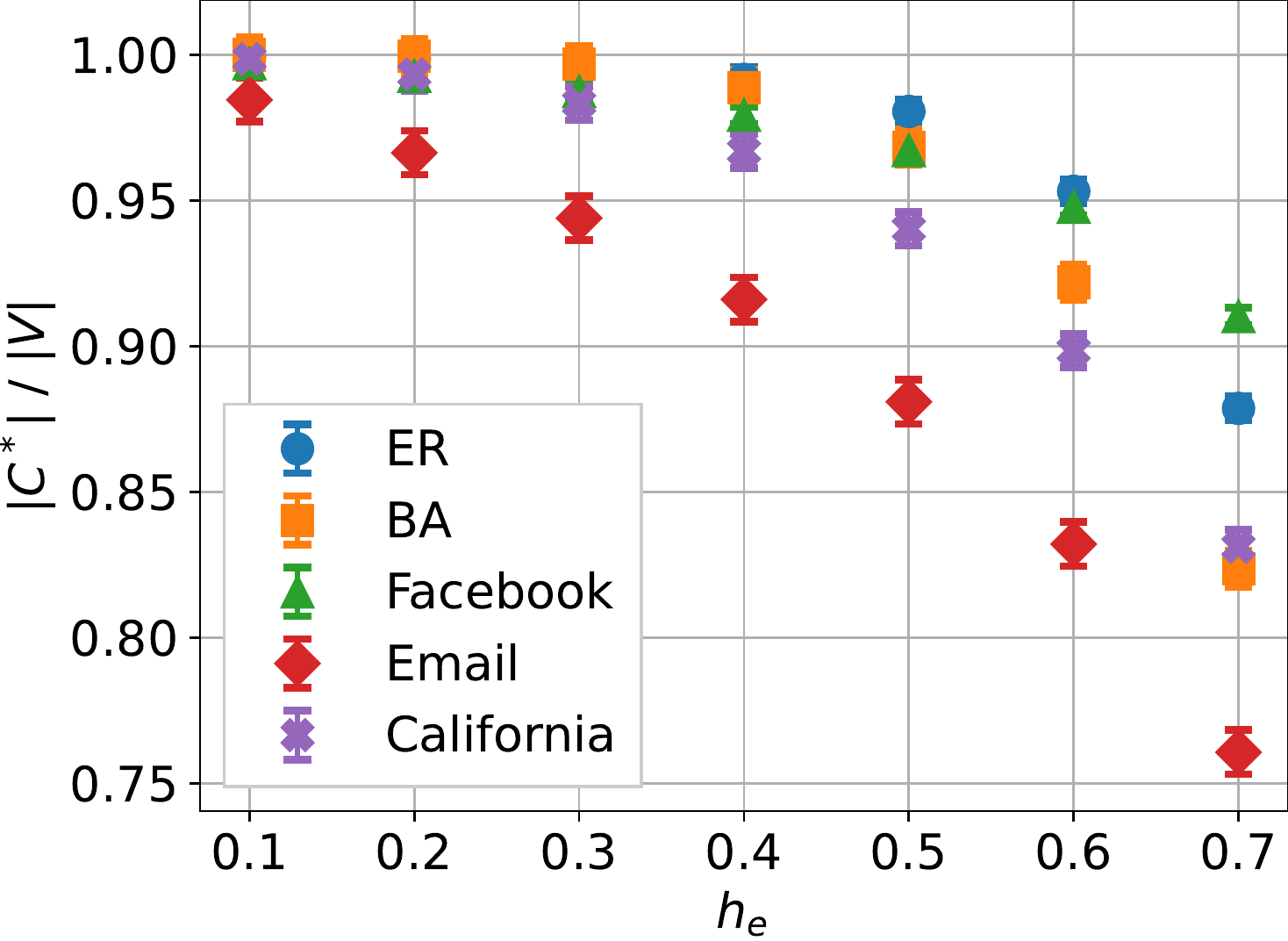}
    \caption{Normalized size of the percolation cluster (largest connected component).
    Each point is an average over $10^3$ iterations.
    Error bars indicate three standard deviations of the mean.
    }
    \label{fig:largest_component}
\end{figure}

On the other hand, examining the size of the percolation cluster (see Fig. \ref{fig:largest_component}) it is clear that it still contains more than 75\% of the modified network, even for the highest considered $h_e$.
This supports the notion that it is the only suitable candidate for running localization algorithms.

While the percolation cluster encapsulates most of the modified network, it may be beneficial to examine the impact of cropping the graph. 
One may consider a link hiding scheme in which $C^* = V$ is guaranteed, i.e. randomly chosen links are hidden only if their deletion would not split the network into multiple components.
We achieve this by checking whether another path $P(v_i, v_j)$ exists after removing link $e_{ij}$.
If such a path exists, the link is considered ``non-essential'' -- its removal did not split the network.
If such a path does not exists then $e_{ij}$ is reinstated and another link is hidden instead.

\begin{figure}[htb]
    \centering
    \includegraphics[width=0.4\columnwidth]{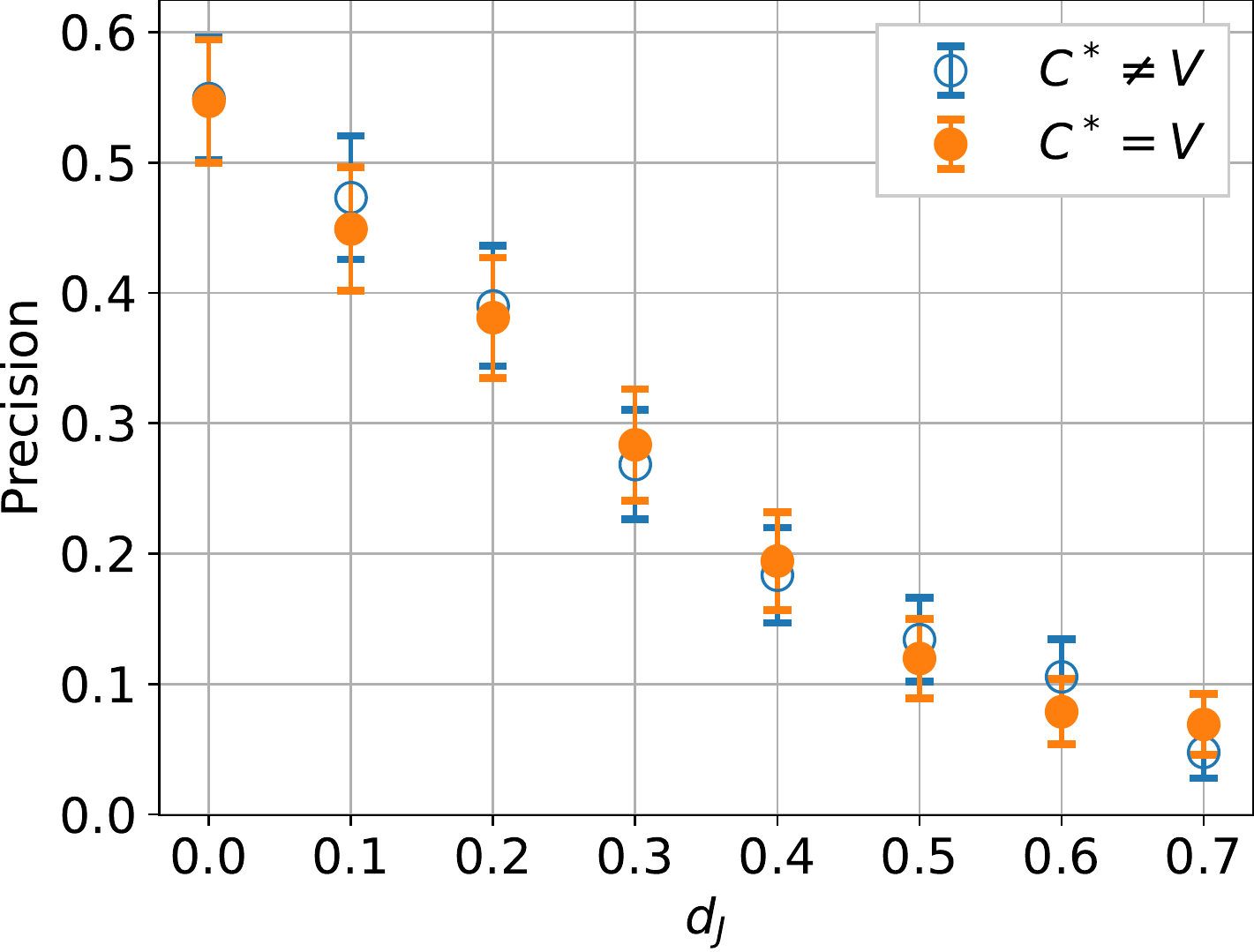}
    \caption{Precision of Pearson algorithm for ER network, modified by hiding $h_e \cdot |E| = d_J \cdot |E|$ random edges.
    Filled markers represent a link hiding scheme in which the network is artificially held together by removing only ``non-essential'' links.
    Parameters $\beta = 0.5$, $r = 0.2$.
    Each point is an average of over $10^3$ iterations.
    Error bars indicate three standard deviations of the mean.
    }
    \label{fig:whole_not_whole}
\end{figure}

This modification to the hiding scheme turns out to have a negligible impact on localization performance.
A comparison of the two hiding schemes has been illustrated in Fig. \ref{fig:whole_not_whole}.
Despite being the most glaring out of all considered parameter/network/algorithm combinations, the difference does not exceed statistical uncertainty.

\section{Suplemental material}
\label{sec:suplemental}
Results presented in Section \ref{sec:results} constitute parameter combinations most representative for the purposes of this paper.
Nevertheless, the output of identical simulations performed for a wider range of $r$ and $\beta$ may be of interest.
Variants of Fig. \ref{fig:results_prec} and Fig. \ref{fig:results_css} are included below as a supplemental material for this exact purpose.

\figpage{0.1}{0.5}
\figpage{0.1}{0.75}
\figpage{0.1}{0.95}

\figpage{0.2}{0.5}
\figpage{0.2}{0.75}
\figpage{0.2}{0.95}

\figpage{0.3}{0.5}
\figpage{0.3}{0.75}
\figpage{0.3}{0.95}

\section*{Acknowledgments}
Research was funded by POB Cybersecurity and data analysis of Warsaw University of Technology within the Excellence Initiative: Research University (IDUB) programme.

\bibliographystyle{model1-num-names}
\bibliography{cas-refs}

\end{document}